\documentclass{aa}
\usepackage{epsfig}

\begin{document}

\title{SPI/INTEGRAL observation of the Cygnus region}

\author{L. Bouchet \inst{1}, E. Jourdain \inst{1}, J. P. Roques \inst{1}, P. Mandrou \inst{1}, P. von Ballmoos \inst{1}, S. Boggs \inst{8}, P. Caraveo \inst{4}, M. Cass\'e \inst{3},  B. Cordier \inst{3}, R. Diehl \inst{2}, P. Durouchoux \inst{3}, A. von Kienlin \inst{2}, J. Knodlseder \inst{1}, P. Jean \inst{1}, P. Leleux \inst{5}, G. G. Lichti \inst{2}, J. Matteson \inst{7}, F. Sanchez \inst{5}, S. Schanne \inst{3}, V. Schoenfelder \inst{2}, G. Skinner \inst{1}, A. Strong \inst{2}, B. Teegarden \inst{9}, G. Vedrenne \inst{1}
\and 
C. Wunderer \inst{2} }

\offprints{L. Bouchet ; {\em bouchet@cesr.fr} }

\institute{
 Centre d'Etude Spatiale des Rayonnements, CNRS/UPS, B.P. 4346, 31028 Toulouse, France
\and
 Max-Planck-Institut für Extraterrestrische Physik, Postfach 1603, 85740 Garching, Germany
\and
DSM/DAPNIA/Sap, CEA-Saclay, 91191 Gif-sur-Yvette, France
\and
IASF, via Bassini 15, 20133 Milano, Italy
\and
 IFIC, University of Valencia, 50 avenida Dr. Moliner, 46100 Burjassot, Spain
\and
Institut de Physique Nucl\'eaire, Universit\'e catholique de Louvain, B-1348 Louvain-La Neuve, Belgium
\and
UCSD/CASS, 9500 Gilman Drive, La Jolla, CA 92093-0111, USA
\and
Space Science Laboratory, University of California, Berkeley, CA 94720, USA
\and
Laboratory for High Energy Astrophysics, NASA/Goddard Space Flight Center, Greenbelt, MD 20771, USA
                }

\date{Received July 17, 2003; accepted August 27, 2003}

\authorrunning{L.~Bouchet}

\abstract{
We present the analysis of the first observations of the Cygnus region by the SPI spectrometer onboard the Integral Gamma Ray Observatory, encompassing ${\sim}$ 600 ks of data. Three sources namely Cyg X-1, Cyg X-3 and EXO 2030+375 were clearly detected. Our data illustrate  the temporal variability of Cyg X-1 in the energy range from 20 keV to 300 keV. The spectral analysis shows a remarkable stability of the Cyg X-1 spectra when averaged over one day timescale.

The other goal of these observations is SPI inflight calibration and performance verification. The latest objective has been achieved as demonstrated by the results presented in this paper.

}

\maketitle

\section{Introduction}
The spectrometer for INTEGRAL (SPI) is one of the two main instruments onboard 
ESA's INTEGRAL observatory launched from Baïkonour, Kasakhstan, on October 17, 
2002. The aim of SPI is to perform  the high resolution spectroscopy as well as 
imaging of astrophysical sources in the energy range between 20 keV and 8 MeV. The imager IBIS (15 keV-10 MeV) (Ubertini et al., 1997) is complementary to SPI with its high angular resolution. JEM-X, an X-ray monitor (Westergaard et al., 1997), and OMC, an optical monitor camera (Gimenez et al., 1997), complete the INTEGRAL observatory.

The INTEGRAL observations of the Cygnus region started on 2002, November, 15 and lasted up to 2002, December, 24. Three high energy sources, namely Cyg X-1,
Cyg X-3 and EXO 2030+375 were clearly detected by SPI.

Cyg X-1  is one of the brightest  and most extensively studied gamma-ray sources in the sky. The system consists of a blue supergiant (HDE 226868) and a 
compact companion generally believed to be the prototype of a black hole candidate. Its high energy emission is characterised by a variability on time scale ranging from milliseconds to months.   
Cyg X-3 is a bright high-mass X-ray binary with a 4.8 hr orbital period,
lying at a distance of ${\sim}$ 10 kpc (Dickey, 1983). The exact nature 
(black hole or neutron star) of the compact object is still in question. 
The companion is most likely an early Wolf-Rayet star (Van Kerkwijk et al., 1992).
The source is highly variable from radio to hard X-rays. It is one of the dozen Galactic binaries considered as a ``microquasar'' (Distefano et al., 2002). 

EXO 2030+375 is a transient accreting X-ray pulsar discovered by EXOSAT 
during a giant outburst in 1985 (Parmar et al., 1989).  Localised at 
$\sim 3.6^\circ$ from Cyg X-3, it is a high mass binary system consisting of
a neutron star and a Be companion. Its X-ray emission is characterised by the 41.8  s neutron star pulsation coupled with recurrent outbursts occurring every 42 days and lasting 7 to 9 days, thought to mark the periastron passage of 
the neutron star. The most extensive observations were made by BATSE (Laycock et al., 2003) while a multiwavelength study was performed by Wilson et al. (2002).

\section{Instrument}
\begin{table*}
\centering
\caption[]{INTEGRAL observations of the Cygnus region 2002 November-December}
\label{tbl-1}
\begin{tabular}{ccccc}
\hline
\noalign{\smallskip}
Revolution & Pointing & Date start - end     &  Accumulation & Useful duration \\
number     & mode     & Year/month/day - hh:mm &   mode        & kilosecond
 \\
\noalign{\smallskip}
\hline
\noalign{\smallskip}
11 & Staring &  02/11/15 - 07:17 - 02/11/18 - 07:02 & photon - photon & 27.8  \\
16 & Staring &  02/11/30 - 06:08 - 02/12/03 - 05:54 & onboard spectra & 56.9  \\
17 & Staring &  02/12/03 - 05:54 - 02/12/06 - 05:39 & photon-photon   & 41.5  \\
18 & Staring &  02/12/06 - 05:39 - 02/12/09 - 05:24 & onboard spectra & 47.4  \\
19 & Dithering & 02/12/09 - 05:24 - 02/12/12 - 05:11 & onboard spectra & 75.7 \\
20 & Dithering & 02/12/12 - 05:11 - 02/12/15 - 05:00 & onboard spectra & 48.5 \\
21 & Dithering & 02/12/15 - 05:00 - 02/12/18 - 04:48 & onboard spectra & 58.0 \\
{} & Dithering & 02/12/15 - 05:00 - 02/12/18 - 04:48 & photon-photon   & 39.3 \\
22 & Dithering & 02/12/18 - 04:48 - 02/12/21 - 04:34 & onboard spectra &123.2 \\
23 & Dithering & 02/12/21 - 04:34 - 02/12/24 - 04:18 & onboard spectra & 75.8 \\
\hline
\end{tabular}
\end{table*}

The SPI spectrometer consists of an array of 19 actively cooled high resolution 
Ge detectors with an area of 508 cm$^{2}$ and a thickness of 7 cm. It is surrounded by a 5 cm thick BGO shield. The detectors cover an energy range of 20 keV - 8 MeV with an energy resolution (FWHM) ranging from  2 to 8 keV as a funtion of energy. In addition to its spectroscopic capability, SPI can image the sky with a spatial resolution of about  $2.6^\circ$ over a field of view of $30^\circ$, thanks to its coded mask. Despite such a modest angular resolution, it is possible to locate sources within few arcminutes. See  Vedrenne et al. (this issue) for more details.

A complete INTEGRAL orbit lasts ${\sim}$ 3 days, but scientific data cannot
be accumulated when the instrument is crossing the radiation belts, reducing the useful time by 30${\%}$. A detailed description of the instrument in-flight performances is given in Roques et al. (this issue). 

\section{Observations}

Each orbit consists in several exposures (typically, corresponding to
the observation of a given source). Owing to the small number of detectors, imaging relies on the possibility of performing the observations in dithering mode (Vedrenne et al., this issue). The pointing direction varies in step of $2^\circ$ with a $5 \times 5$ quadratic or 7 hexagonal pattern within an exposure. Each pointing has  a duration generally fixed to $\sim$ 30 minutes. 

Observations can be performed either in spectral mode, where the single events 
are accumulated  in one unique spectrum for each pointing, or in photon-photon mode, where all events are time tagged with a temporal resolution of $102.4 \mu$s, and stored individually.

The data used in this analysis were recorded from 2002 November 15 to 2002, December 24 (table 1). Variations of the temperature of the Ge detectors produce a small gain drift in their associated electronics. Such an effect is easy to correct revolution per revolution, through the energy calibration of each detector using known background induced lines. For the analysis described in this paper, we used only  the single detector events (see Diehl et al., this issue).

\section{Data analysis}

\subsection{Images analysis}
\subsubsection{Misalignment between SPI  and the star trackers}
The first images  of both  Cyg X-1 and the Crab nebula yielded positions on average 8 arcmin away from their correct values. 
Such a misalignment between the star trackers and the SPI axes had been measured on ground and later found stable after launch. A correction has thus been computed and included in the  data analysis. This correction is similar to that obtained by Walter et al. (this issue), which was not available at the time of this analysis.

\subsubsection{Background determination}
 
The background substraction method depends on the pointing mode. For exposures in dithering mode, when a bright source illuminates a different combination of detectors in each step of the dithering pattern, it is possible to determine the background level for each detector independently.  
In the staring mode, the unique pointing direction makes it impossible to determine background mathematically. A model of the background distributions 
on the detector plane has to be assumed. Analyses of ``Empty-field'' observations
have shown that, for a given energy, the relative count rates of the 19 Ge detectors are broadly constant while the global amplitude (normalisation factor) varies with time. We thus fix the relative count-rate distribution  (depending on the energy band) leaving the normalisation as the only free parameter.  

To  limit the number of free parameters, we fix the background values over 
a timescale as long as possible. Considering the  flux variations measured 
by the Anticoincidence Shield or Ge saturated events, we assume that the background parameters remain constant over typically half an orbit, in both staring and  dithering modes.

\subsubsection{Images}
We have used an iterative source removal  algorithm to build sky images (SPIROS, Skinner et al., this issue).  It is always difficult to build images with a strong and variable source.  The main uncertainties come  from the background determination. Although the background is generally quite stable, (solar flares 
or  radiation belts regions excepted), variations of several percents of the 
background along with source variability can yield unreliable results. 
In order to limit these effects, we have not used pointings for which reconstructed images (convolved by the reponse matrix), when compared to detector counts, give an unacceptable $\chi^{2}$. Those precautions reduced the number of  pointings used by ${\sim}$ 10 \%. For each image, we are looking for a maximum of 10 sources or all sources above a threshold level of 2 $\sigma$.

Fig. 1 shows an image of the Cygnus region, in the energy range between 20 and 84 keV, obtained by SPI using all the cleaned data, i.e. ${\sim}$ 600 ks. 

\begin{figure}[here]
  \includegraphics[width=8.8cm]{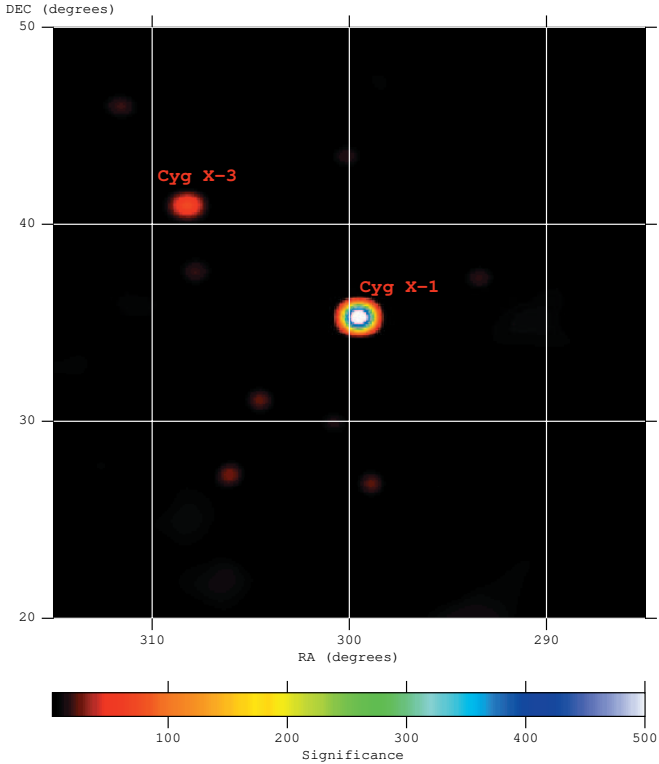}
\caption{An Image of the Cygnus region using all data listed in table 1 between 20 and 84 keV. The significance levels for Cyg X-1 and Cyg X-3 are 684  $\sigma$ and 74  $\sigma$ respectively.}
\end{figure}

It demonstrates, first of all, that our misalignment correction is valid: the SPI position for Cyg X-1 turned out to be $\alpha_{2000}=299.559 \pm 0.016$ $\delta_{2000} =35.202 \pm 0.012$, compatible, within less than 2 arcmin, with its known position. The statistical significance of the detection of Cyg X-1 is ${\sim}$ 680 $\sigma$ (20 - 84 keV).

Fig. 1  shows  also the presence of  Cyg X-3 at $\alpha_{2000}$= 308.222 $\pm$ 0.090 $\delta_{2000}$=40.929 $\pm$ 0.062, compatible with the source theoretical position, with a significance of 74 $\sigma$.

\begin{figure}[tb]
  \includegraphics[width=8.8cm]{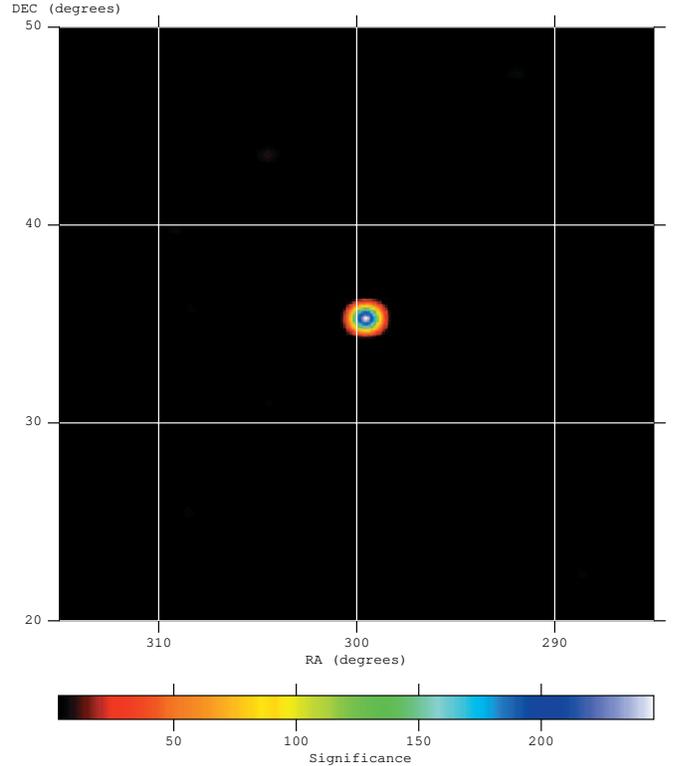}
\caption{An image of the Cygnus region as above between 84 and 308 keV. The significance level for Cyg X-1 is 247  $\sigma$ while Cyg X-3 is not detected.}
\end{figure}

\begin{figure}[tb]
  \includegraphics[width=8.8cm]{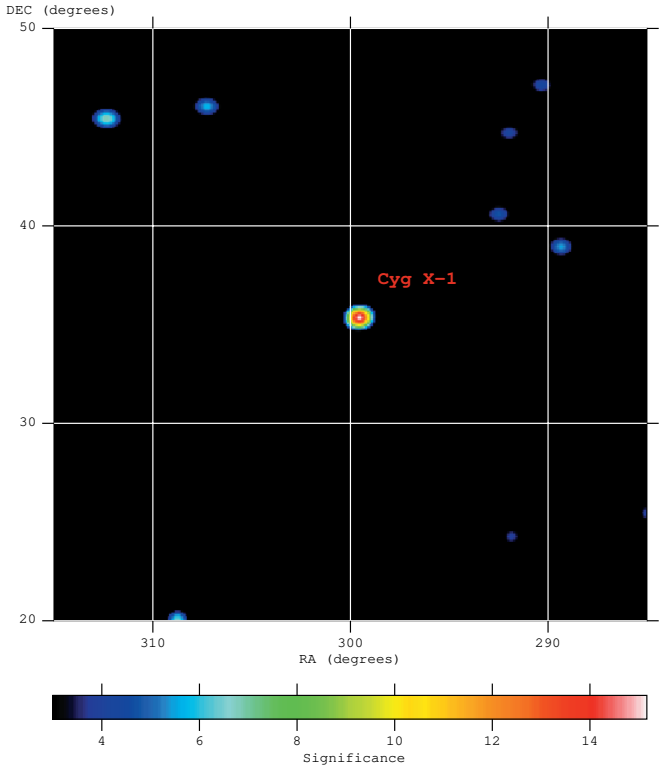}
\caption{An Image of the Cygnus region using all data listed in table 1 between 308 and 615 keV. The significance level for Cyg X-1 is 16  $\sigma$.}
\end{figure}

When selecting higher energy ranges, as in figure 2 and 3, reconstructed using photons in the 84 - 308 keV and 308 - 615 keV energy ranges, only Cyg X-1 continues to be detected.  

Figure 4 contains only the period of activity of EXO 2030+375 , namely 
revolution 19 to 22, with a total of 345 ks of observing time (see section 4.2).

\begin{figure}[tb]
  \includegraphics[width=8.8cm]{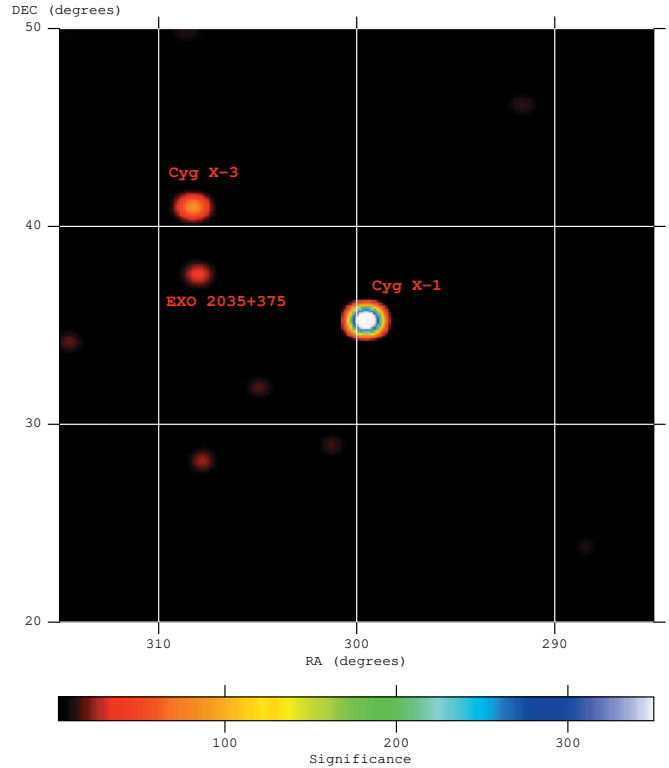}
\caption{Image of Cygnus region for revolutions 19 to 22 (see table 1) in the energy range between 20 and 50 keV.}
\end{figure}

EXO 2030+375 is  found at $\alpha_{2000}$=308.02 $\pm$ 0.10 $\delta_{2000}$=37.55 $\pm$ 0.08, compatible with its known position.

In figure 4 (20 - 50 keV), the statistical significance of sources is as follows:
Cyg X-1 is detected at 570 $\sigma$, Cyg X-3  at 86 $\sigma$ and EXO 2030+375
at 36 $\sigma$.

All the other excesses are found with a significance less than 15 $\sigma$, i.e  
less than 3\% of  the strongest source. The presence of noise at this level can be explained in terms of Cyg X-1 variability, background determination procedure, and/or systematic effects due to uncertainties in the calculated coded mask response.

\subsection{Light curves}
The light curves of the three sources are obtained simultaneously, with the background fixed to the mean orbit values (see section 4.1.2). The source positions are left free but they are required to be constant for the whole set of observations.

For Cyg X-3 and EXO 2030+375, the pointings have been regrouped to get a timescale of ${\sim}$ 5-6 hours.

The light curve of Cyg X-1 (fig 5) shows a marked variability at low energy (20 - 48 keV), by a factor of ${\sim}$ 2.

\begin{figure}[here]
  \includegraphics[width=8.8cm]{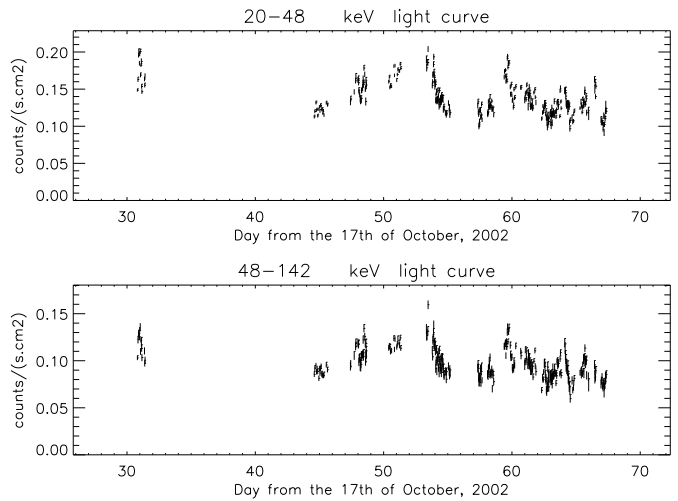}
\caption{The light curve of Cygnus X-1 between 20 and 48 keV and 48 and 142 keV with a time resolution of $\sim$ 30 minutes}
\end{figure}

Fig. 6 and 7 show  the light curves of Cyg X-3 and EXO 2030+375 between 20 and 45 keV. Although the statistics are much less impressive than in the case of Cyg X-1, we can see that Cyg X-3 is detected through all the revolutions, while EXO 2030+375 exhibits temporal evolution corresponding to an expected periastron outburst, with no detection before day 2002/12/09, then a significant emission from  2002/12/10  to   02/12/20 (middle of revolution 19 through revolution 22, see table 1).

\begin{figure}[here]
  \includegraphics[width=8.8cm]{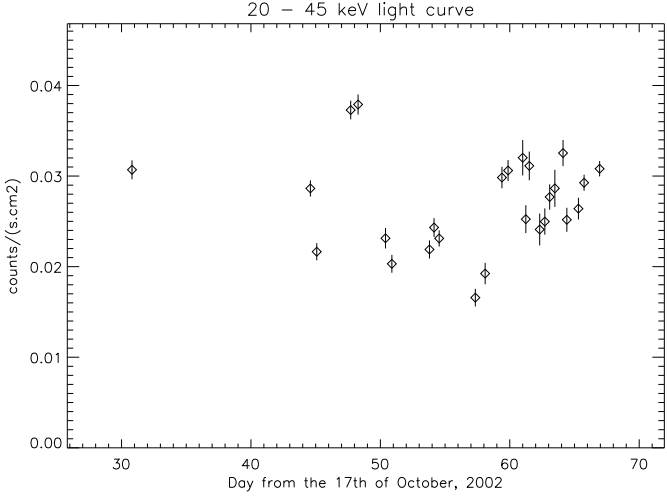}
\caption{Light curves of Cygnus X-3 with a time resolution of 5-6 hours (20-45 keV)}
\end{figure}

\begin{figure}[here]
  \includegraphics[width=8.8cm]{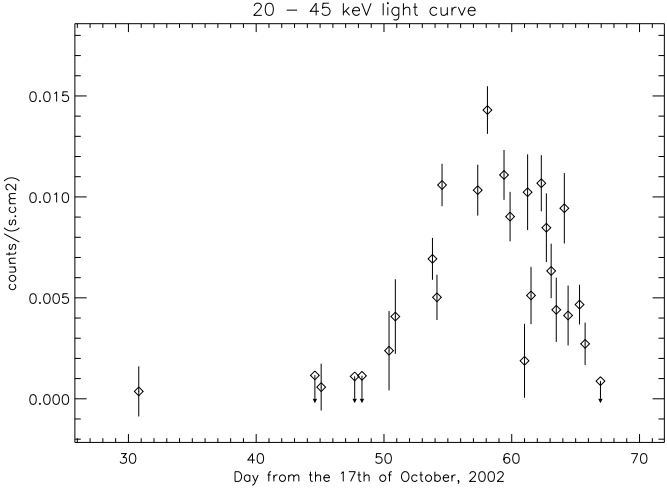}
\caption{Light curves of EXO 2030+375 with a time resolution of 5-6 hours (20-45 keV)}
\end{figure}

\subsection{Spectra}
Counts and photon spectra for all the sources in the field of view are extracted simultaneously by an algorithm which fits all the positions and intensities of a set of given sources. To build  photons spectra, we need the energy response matrix corresponding to each pointing. We used 2003, May version of the energy response matrix (Sturner et al., this issue).

\subsubsection{Cyg X-1}
Fig. 8 shows the  spectrum of Cyg X-1 averaged over the total exposure time of ${\sim}$ 600 ks.

\begin{figure}[here]
  \includegraphics[width=8.8cm]{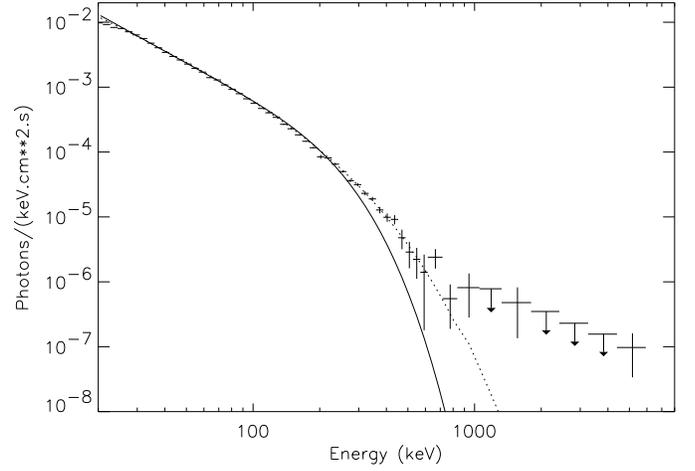}
\caption{Mean photon spectra of Cyg X-1 for the entire set of observations (${\sim}$ 600 ks). Solid line is Sunyaev-Titarchuck (S-T) model and dotted line is the power law + exponential cutoff model. Upper limits are at 1 $\sigma$. }
\end{figure}

The source is detected up to ${\sim}$ 1 MeV and its spectrum is well described, 
at least up to ${\sim}$ 300 keV, by the usual Sunyaev-Titarchuck (S-T) inverse Compton emission model (Sunyaev and Titarchuck, 1980) or by a power law with an exponential cutoff. For a spherical geometry, the temperature (kT$_e$) is ${\sim}$ 48 keV and the optical depth  is $\tau$ ${\sim}$ 2.5 for the S-T model while for the second model the power law index is ${\sim}$ 1.5 and the cutoff energy ${\sim}$ 155 keV. 
A thermal law gives a temperature of ${\sim}$ 70 keV. In oder to compare our results with those obtained by OSSE/GRO, where the fitting was performed 
between 60 keV and 4 MeV (Phlips et al., 1996), we computed also the best fitting parameters above 60 keV. The new fit yielded higher kT$_e$ ${\sim}$ 58 keV with a lower optical depth $\tau$ ${\sim}$ 2.0, a set of parameters fully compatible with those of OSSE/CGRO.

Above 300 keV, we detect an excess with respect to the S-T model, similar to that measured by OSSE/GRO (Phlips et al., 1996). We cannot  exclude that the variability of the source produces such a feature when the data are integrated over a long time.

The individual spectra, built on timescale of one orbit, reflect the variability of the source intensity, but the spectral parameters remain remarkably stable, with $kT_e$ varying from 45 to 54 keV  and $\tau$ from 2.2 to 2.6,  the power law with an exponential cutoff model gives $\alpha$ between 1.5 and 1.7 and a cutoff energy in the energy range 144 - 205 keV. We have to note that the $\chi^{2}$, obviously rather high, are lower with the exponential cutoff power law model. 

\subsubsection{Cyg X-3}
The Cyg X-3 average spectrum is shown on fig. 9. 

\begin{figure}[here]
  \includegraphics[width=8.8cm]{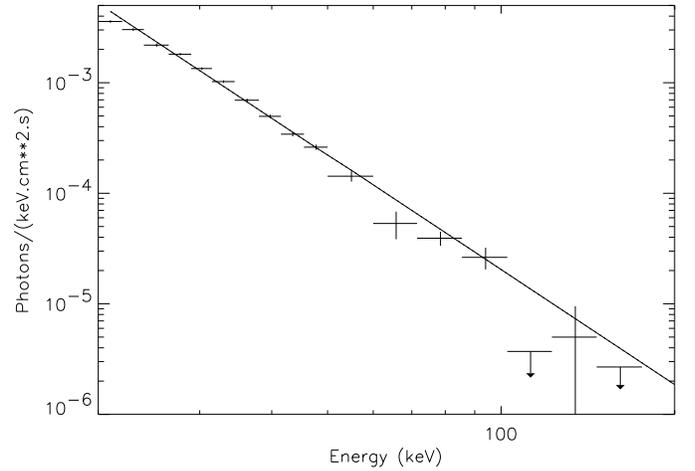}
\caption{The average spectrum of Cyg X-3 for a total exposition of 600 ks. The solid line is a best fit power law model, with an index of 3.45 $\pm$ 0.01 and a flux at 50 keV of 2.19 $\pm$ 0.01 10$^{-4}$ photons$\cdot$cm$^{-2}\cdot$s$^{-1} \cdot$keV$^{-1}$.
Upper limits are at 1 $\sigma$.}
\end{figure}

Cyg X-3 is detected up to ${\sim}$100 keV and its average photon spectrum is well fitted by a power law of index ${\sim} $3.45 with a flux at 50 keV of ${\sim} 2.2  \times 10^{-4}$ photons$\cdot$cm$^{-2}\cdot$s$^{-1}\cdot$keV$^{-1}$.
This index value is similar to that reported by OSSE (McCollough et al., 1999), but softer than the non-thermal power law component index derived by Choudhury et al.(2002) with ASM+BATSE data.

\subsubsection{EXO2030+375}

\begin{figure}[here]
  \includegraphics[width=8.8cm]{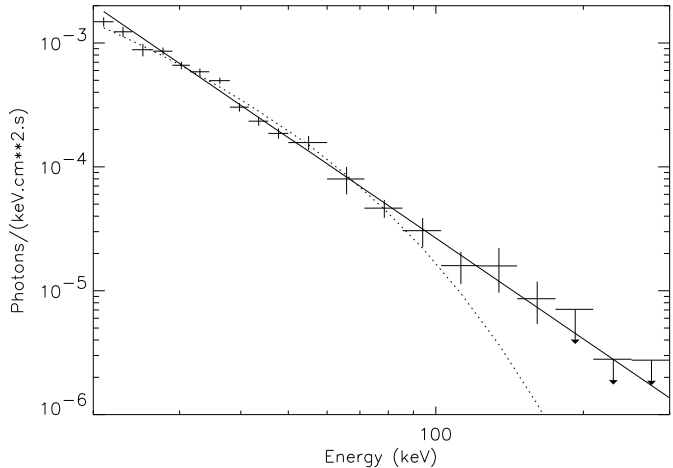}
\caption{The average spectrum of EXO 2030+375 for a total exposition time of 345 ks.
The best fit power law (solid line) has an index of 2.72 $\pm$ 0.02 and the best fit thermal model (dotted line) has a temperature of 27.2 $\pm$ 0.1.
Upper limits are at 1 $\sigma$.}
\end{figure}

EXO 2030+375 is also detected up to 150 keV (fig. 10). Its average photon spectrum (revolution 19 to 22, 345 ks exposure) is well fitted by a thermal model with temperature ${\sim}$ 27 keV  or  a power law of  index
${\sim}$ 2.7.

A possible spectral feature at ${\sim}$ 36 keV, tentatively explained as a cyclotron absorption line, has been suggested by Reig and Coe (1999).

We do not detect this feature in the averaged spectrum with a 2 $\sigma$ upper limit between 34 and 38 keV of $ 8 \times 10^{-5}$ ph$\cdot$cm$^{-2}\cdot$ s$^{-1}$. We cannot exclude the presence of such a feature in shorter timescales, but this requires further analysis. 

\section{Discussion and Conclusion}
The emission from Cyg X-1  has been detected up to 1 MeV. While, in the energy range between 20 and 300 keV, its flux varies by a factor of ${\sim}$ 2, its spectral shape remains remarkably stable, on a one day timescale. Our mean photon spectrum is compatible with the average CGRO/OSSE spectrum in the hard state (McConnell et al., 2002).

An outburst of EXO 2030+375 occured during our observations. The average SPI spectrum during this  6-7 days outburst is compatible with that obtained by BATSE (Wilson et al, 2002).

Initial results obtained on Cyg X-1, Cyg X-3 and EXO 2030+375 are coherent with the knowledge we have on those sources and allow us to proceed with a more detailled analysis.

Note that, although the angular distance between Cyg X-3 and EXO 2030+375 is close to the angular resolution of SPI,  we can easily measure their fluxes and 
positions.

Cyg X-1 is one of the strongest persistent source in the SPI/INTEGRAL energy domain. 
From a technical point of view, it is used to test the coherence of our data 
analysis system, performance verification and inflight calibrations. Those objectives have been achieved and they demonstrate that SPI is working perfectly. First of all, at the present time, the quality of the images leads us to conclude that the  spatial and energy responses are accurate to about 2-3 \%.

While qualitative results will not change, quantitative results are subject to
the incertainties in the used energy response. We used the  May 2003 version of the response matrix (Sturner et al., this issue), which underestimates the efficiency at low energy, resulting in steeper spectra.

We plan to improve the data analysis system by using more precise background analysis and modelling and by including source variability in the image analysis package.

\begin{acknowledgements}
  The SPI/INTEGRAL project has been completed under the responsibility and leadership of CNES. We are grateful to ASI, CEA, CNES, DLR, ESA, INTA, NASA and OSTC for support.
\end{acknowledgements}

\end{document}